\documentclass[a4paper]{article}
\usepackage{ifthen}
\usepackage{epsfig}
\usepackage{amssymb}
\usepackage{graphicx}
\usepackage{amsmath}
\usepackage[center]{subfigure}
\newcommand{\ba}{\begin{eqnarray}}
\newcommand{\ea}{\end{eqnarray}}
\newcommand{\be}{\begin{equation}}
\newcommand{\ee}{\end{equation}}
\newcommand{\Ds}{D_{(s)}}
\newcommand{\MeV}{\mbox{MeV}}
\newcommand{\GeV}{\mbox{GeV}}
\begin{document}
\begin{titlepage}
\vfill
\begin{center}
{\Large\bf Upper bounds on $f_D$ and $f_{D_s}$\\ 
from two-point correlation function in QCD}\\[1.6cm]
{\large\bf
Alexander~Khodjamirian}\\[0.5cm]
{\it  Theoretische Physik 1, Fachbereich Physik,
Universit\"at Siegen,\\ D-57068 Siegen, Germany }\\      
\end{center}
\vfill
\begin{abstract}
The correlation function 
of two pseudoscalar charmed quark currents 
with a positive hadronic spectral density
is employed to obtain upper bounds
on the decay constants of $D$ and $D_s$ mesons.
Including all known terms of the 
operator-product-expansion of this correlation function
in QCD
and taking into account the estimated uncertainties, 
we obtain  $f_{D}<230 $ MeV and $f_{D_s}<270$ MeV.
Comparison with the 
decay constants determined from  $D\to l \nu_l$ 
and $D_s\to l \nu_l$ measurements,
reveals a tension between  the bound 
and current experimental value of $f_{D_s}$.

\end{abstract}
\vfill
\end{titlepage}

\section{Introduction}
  
The decay constants of charmed $D^+$ and $D_s$ mesons, 
defined via the hadronic matrix elements:
\begin{equation}
\langle 0|\bar{d} \gamma_\mu\gamma_5 c |D^+(p)\rangle =
if_{D}p_\mu\,,~~
\langle 0|\bar{s} \gamma_\mu\gamma_5 c |D_s(p)\rangle =
if_{D_s}p_\mu\,,
\label{eq:def}
\end{equation} 
are extracted from the branching fractions of 
purely  leptonic 
decays $D^+\to l^+ \nu_l$ and $D_s\to l^+ \nu_l$ ($l=\mu,\tau$),
respectively. Recent CLEO measurements of these decays yield
\cite{Zhang,Stone}:   
\ba
f_{D}&=& 205.8\pm 8.5\pm 2.5~\MeV,~~~\label{eq:expD}  \\
f_{D_{s}}&=&267.9\pm 8.2\pm 3.9 ~\MeV\,,
\label{eq:expDs}
\ea
assuming $|V_{cd}|=|V_{us}|$ and $|V_{cs}|=|V_{ud}|$.
Other experimental results for $f_{D}$ and $f_{D_s}$ are 
consistent with the above intervals (see \cite{RosnerStone} 
for a review).  
Note that the observed $SU(3)_{fl}$-violation in
$D_s$ and $D$ decay constants 
turns out to be larger than in the light pseudoscalar 
mesons where $f_K=155.5$ MeV and $f_\pi=130.4$ MeV \cite{PDG}. 

The decay constants of heavy mesons are accessible in 
lattice QCD. The recent results: 
$ f_D= 207\pm 4$ MeV and  $f_{D_s}= 241\pm 3$ MeV \cite{Follana},
obtained with the number of sea-quark flavours 
$N_f=3$ have quite small errors.
The lattice value of $ f_D$ is  in a good 
agreement with (\ref{eq:expD}), whereas
$f_{D_s}$ is smaller than (\ref{eq:expDs}). This puzzling situation  
caused discussions of a possible 
non-standard physics (see e.g., \cite{Dobrescu,NarisonV}). 
The previous $N_f=3$  lattice QCD result \cite{Aubin}:
$ f_D= 201\pm 3\pm 17$ MeV, $f_{D_s}= 249\pm 3\pm 16$ MeV,
as well as the two recent $N_f=2$ calculations: 
$f_D= 205\pm 7\pm 7$ MeV, $f_{D_s}= 248\pm 3\pm 8$ MeV  
\cite{Blossier} and 
$f_{D_s}= 257\pm 3\pm 3\pm 5$ MeV (preliminary) 
\cite{vonHippel}  quote larger 
errors,  still their central values for $f_{D_s}$ 
are smaller than in (\ref{eq:expDs}). Recent results for
$f_{\Ds}$ in quenched lattice  QCD can be found in \cite{Ali Khan,Heitger}
and a discussion of the accuracy of lattice determinations 
in \cite{Heitger}.

An alternative way 
to calculate $f_{\Ds}$ is provided by QCD sum rules \cite{SVZ}.
The method is based on the operator-product 
expansion (OPE) of the two-point 
correlation function evaluated in deep 
spacelike region, in terms of perturbatively 
calculable Wilson coefficients and QCD vacuum condensates.
The result of OPE 
is related with the hadronic
matrix element (\ref{eq:def})  via dispersion relation 
and quark-hadron duality.
The QCD sum rule calculation of the heavy-meson 
($\Ds$ and $B_{(s)}$) decay constants has a long history, starting from \cite{6auth,AE,Broadhurst}. More  recent calculations 
using QCD sum rules are in \cite{JL,PS,Narison} and 
finite-energy sum rules in \cite{Bordes}.
A review of earlier results can be found
in \cite{CK}.
Not going into further details, let us only mention
that QCD sum rules yield
$f_{D_s}>f_D$, and the predicted intervals for 
both decay constants 
are in the ballpark of the lattice QCD results,
albeit with larger uncertainties.
In particular, a ``systematic'' uncertainty of 
QCD sum rules which is difficult to quantify 
in a model-independent way, is caused by the quark-hadron 
duality approximation.  
Having in mind a possible confrontation 
with experiment, an update of 
the QCD sum rule predictions for $f_{\Ds}$ is 
a timely task which is however 
beyond our scope here.

The aim of this paper is to remind  
that the same correlation function which is used to 
obtain the QCD sum rule, 
provides upper bounds  on $f_D$ and $f_{D_s}$. These bounds
simply follow from the positivity of the 
hadronic spectral density of the correlation function
and are independent of 
the quark-hadron duality  approximation.
In what follows, we will calculate 
the upper bounds for $D$ and $D_s$ meson decay constants 
and find that the bound on $f_{D_s}$ is on the verge 
of disagreement with the experimental value.   

\section{Derivation of the bounds} 
We start from the correlation function of two  
charmed pseudoscalar quark currents 
$j_5=(m_c+m_{d}){\bar{d}}i\gamma_5 c$, 
the divergence of the axial-vector current 
in (\ref{eq:def}): 
\begin{eqnarray}
\Pi(q^2)=i \int d^4xe^{iqx}
\langle 0\mid T\{j_5(x)j_5^\dagger(0) \}\mid 0\rangle
=\sum\limits_{h=D, D^*\pi,...}
\frac{\langle 0\mid j_5 |h\rangle\langle h | j_5^\dagger\!\mid 0\rangle}{m_h^2-q^2}\,.
\label{eq:corr}
\end{eqnarray}
For definiteness, the channel of the charged $D$ meson is 
considered, the corresponding
correlation function $\Pi_s(q^2)$ 
for $D_s$-channel is obtained by a simple replacement 
of  the light-quark  flavour $d\to s$. On the r.h.s. of
(\ref{eq:corr}), 
the correlation function, via unitarity condition, is written 
as a sum over all hadronic states with $D$ quantum numbers, 
a very schematic representation of the 
dispersion integral
over hadronic spectral density. The ground $D$-state 
contribution to (\ref{eq:corr}) contains the square 
of the decay constant: 
$\langle 0| j_5|D\rangle =f_{D}m_{D}^2$.  

At large virtualities, $q^2\ll m_c^2$, 
the correlation function (\ref{eq:corr}) 
is dominated by short distances. In that region  
we approximate $\Pi(q^2)$ by 
OPE, including  the contributions of 
the perturbative quark-loop diagrams 
(with gluon radiative corrections) and the 
terms with vacuum condensates. The latter are 
suppressed by inverse
powers of $m_c^2-q^2$ and it is sufficient to
include condensates up to dimension $d\leq 6$. 
For virtual $c$ and light quarks 
propagating in the correlation function, it 
is convenient to adopt  the $\overline{MS}$ scheme. The 
most complete  expressions  for OPE in this scheme 
are presented in \cite{JL} where they were 
used for the QCD sum rule calculation of 
$f_B$ and $f_{B_s}$. The corresponding 
expressions for $\Ds$-meson case
are simply obtained by replacing the heavy quark mass 
$m_b\to m_c$, hence, there is no need to 
represent here the explicit formulae. We only write  
down a schematic form of the OPE result after 
Borel transformation: 
\begin{eqnarray}
\Pi(M^2)= \sum\limits_{n=0,1,2}\,\,\int\limits_{(m_c+m_d)^2}^\infty \!\!ds 
\left(\frac{\alpha_s}{\pi}\right)^n\rho^{(n)}(s)e^{-s/M^2}
\nonumber \\+
 \sum\limits_{n=0,1}\left(\frac{\alpha_s}{\pi}\right)^n
\Pi^{(n)}_{\langle \bar{q}q\rangle }(M^2)+
\sum\limits_{d=4,5,6}\Pi_{d}(M^2)\,.
\label{eq:OPE}
\end{eqnarray}
The Borel parameter $M^2$ replaces $q^2$ and represents
the virtuality scale at which the correlation function
is calculated. In the above, the perturbative contributions  
of quark-loop diagrams are represented
in a convenient form of dispersion integrals over 
spectral densities, 
$\rho(s)=\frac{1}{\pi}\mbox{Im}\Pi(s)$. The 
$\alpha_s$-expansion of $\rho(s)$  
includes the two-loop \cite{Broadhurst} and three-loop 
\cite{ChetyrkinStein} gluon radiative corrections. 
For the latter the program 
'Rvs.m', kindly  provided by the authors 
of \cite{ChetyrkinStein} 
was used. The second line in (\ref{eq:OPE}) 
includes the contributions of the quark condensate
with the $O(\alpha_s)$ correction calculated 
in \cite{JL}, as well as  
the standard gluon-, quark-gluon- and 4-quark-condensate 
terms  of OPE denoted by the respective condensate dimension 
$d=4,5,6$.
   
Borel-transforming (\ref{eq:corr}), we use the positivity of the 
hadronic spectral density. Hence, $\Pi_{(s)}(M^2)$ provides
an upper limit for the ground-state $\Ds$-meson contribution 
to the hadronic sum in (\ref{eq:corr}). The resulting upper bound
for the $\Ds$ decay constant is:~\footnote{Positivity bounds of various form 
for this correlation function can be found in 
the literature starting from \cite{Narison1}.} 
\be
f_{\Ds}< \sqrt{\Pi_{(s)}(M^2)
\frac{e^{m_{\Ds}^2/M^2}}{m_{\Ds}^4}}\,.
\label{eq:bounds}
\ee
where $\Pi_{(s)}(M^2)$ is calculated from (\ref{eq:OPE}). 
The bound is valid at any $M^2$,
at which one can trust the OPE (\ref{eq:OPE}). 
Naturally, one has to find  the most restrictive 
value of the r.h.s. in (\ref{eq:bounds}).

\section{Numerical analysis} 

Numerical results for the bounds 
are obtained adopting
the  $c$-quark mass interval
$\bar{m}_c(\bar{m}_c)=1.29\pm 0.03$ GeV which
covers the recent determinations \cite{mc} 
from charmonium sum rules with $O(\alpha_s^3)$ accuracy, 
and, conservatively, we double the uncertainty.       
Note a good agreement of this determination with 
the recent lattice QCD result 
\cite{Allison} for $\bar{m}_c$.
For the strange quark mass we adopt  
$m_s(2 \,\mbox{GeV}) = 98\pm 16 $ MeV, 
the interval of 
QCD sum rule determinations with $O(\alpha_s^4)$ accuracy
\cite{ms}. Fixing $m_s$ and using the 
ChPT relations \cite{Leutwyler}
we obtain  $m_u(2 \mbox{GeV})=2.8\pm 0.6$ MeV and 
$m_d (2 \mbox{GeV})= 5.2\pm 0.9 $ MeV.  
The quark-gluon coupling 
and quark masses are taken with 4-loop running, 
employing the program  provided in \cite{RunDec},
with $\alpha_s(m_Z)=0.1176 \pm 0.002$ \cite{PDG}. 
We assume equal renormalization scales 
for the quark masses and $\alpha_s$.
The quark condensate density is obtained 
using Gell-Mann\,-\,Oakes\,-\,Renner relation: 
$\langle 0| \bar q q |0 \rangle(1~\GeV) = 
-m_\pi^2f_\pi^2/[2(m_u+m_d)(1~\GeV)]=-(250^{+16}_{-12}~\MeV)^3$,
(with $f_{\pi}=(130.4\pm 0.04\pm 0.2)$ MeV and
$m_\pi=139.57$ MeV \cite{PDG}). 

The ratio of the strange and nonstrange quark condensate 
densities $\langle 0|\bar{s}s |0\rangle=
(0.8\pm 0.3)\langle 0| \bar{q}q |0\rangle $,
as well as the intervals for 
the quark-gluon, gluon and four-quark condensate densities
not quoted here for brevity, are taken as in \cite{CKP},
where a different correlation function using 
these universal parameters was calculated
(see also \cite{Ioffe} for a review of 
vacuum condensates). The suppression of $d=4,5,6$
terms in OPE makes uncertainties of the
respective condensate densities 
inessential for our numerical results.
\begin{figure}
\label{fig:uplim}
\begin{center}
\includegraphics[width=6cm]{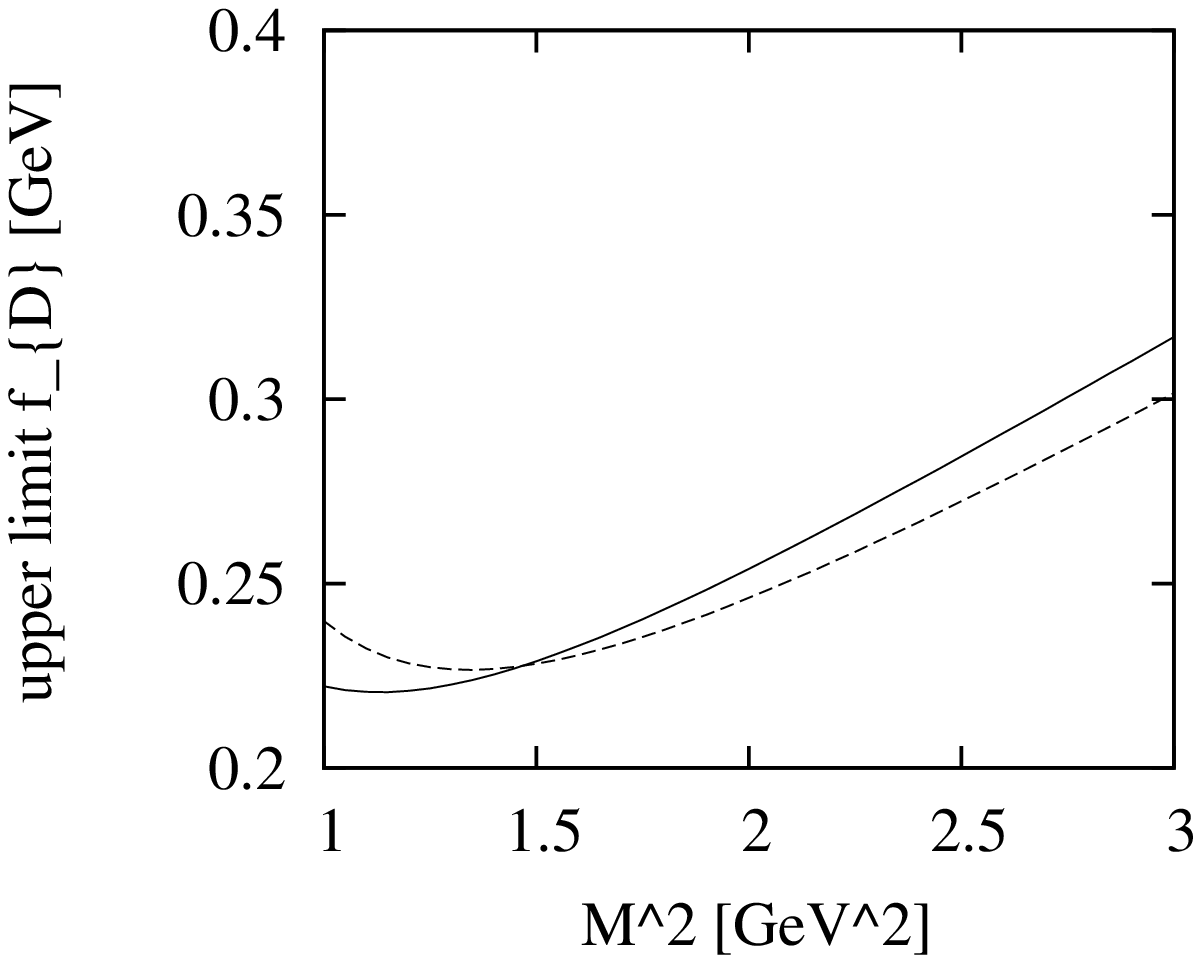}
\includegraphics[width=6cm]{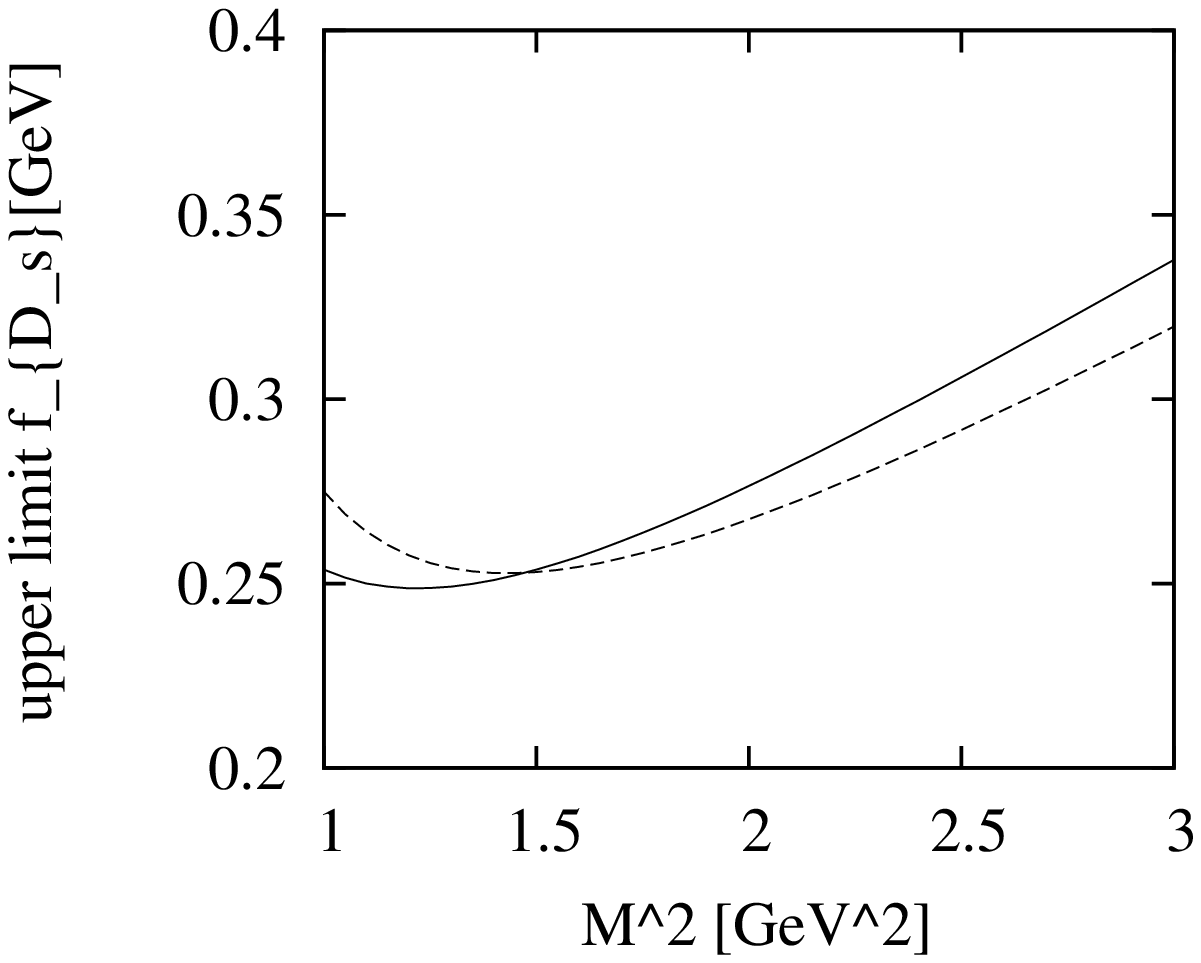}\\
\end{center}
\caption{\it  The upper bounds for $f_D$ (left) 
and for $f_{D_s}$ (right) as functions of the Borel
parameter squared at $\mu=1.5 \GeV$ (solid) and 
$\mu=3.0 \GeV$ (dashed).}
\end{figure}

The upper bounds on $f_D$ and $f_{D_s}$ 
calculated from (\ref{eq:bounds})
are plotted in Fig.1 as a function of 
$M^2$ at central values of all input parameters. 
The default renormalization scale is  
$\mu=1.5$ 
GeV, close to $\mu\sim \sqrt{m_{D}^2-m_c^2}$ used
in sum rule calculations. In $\overline{MS}$ scheme 
the perturbative expansion 
works reasonably well: $O(\alpha_s)$ and 
$O(\alpha_s^2)$ terms in (\ref{eq:OPE}) 
are, respectively, $\leq 30\%$ and $\leq 10 \%$ of the total 
perturbative contribution.  

As expected, the bound for $f_{D_s}$ is larger than the 
one for $f_D$. Both bounds grow and become
less restrictive at larger $M^2$, 
implicitly due to increase of the 
relative weight of the continuum and excited  
states in the Borel-transformed hadronic sum. 
The most restrictive upper bounds 
\be
f_{D}< 220 ~\MeV,~~ f_{D_s}< 250 ~\MeV \,,
\label{eq:1stbound}
\ee
are reached around $M^2=1.2$ GeV$^2$.
At these values of the Borel scale,  the sum of 
power suppressed $d=4,5,6$ condensate contributions
is less than 4\% of the total $\Pi(M^2)$,
hence the condensate expansion can be trusted.
The scale dependence turns out  
to be mild, e.g., increasing the scale from $\mu=1.5$ GeV 
to $\mu=3$ GeV, one obtains the most restrictive bounds at 
$M^2\simeq 1.4$ GeV$^2$, and they are only slightly (by about 
+5 MeV) shifted (see Fig.1). Decreasing the
scale up to $\mu=1.0$ GeV, produces a more pronounced
shift, by about -15 MeV, so that the bounds  
are even more restrictive. 
However, the NNLO correction at this scale reaches 
$\simeq 20 \%$ of the total perturbative contribution, signalling 
that the perturbative expansion  is less convergent 
numerically. Hence, to be on a conservative side, 
we use the bounds obtained at the default scale, including 
their uncertainty (at fixed $M^2$), caused by the variation 
1 GeV $<\mu <3$ GeV, in the error budget discussed below.

The bounds have uncertainties caused by the limited accuracy of 
QCD parameters (quark masses, $\alpha_s$ and condensate
densities) and by the scale-dependence.
The individual uncertainties
are estimated by varying all inputs one-by-one within the adopted 
intervals. The results are collected in the table: 
\vspace{-1mm}
\begin{center}
\begin{tabular}{|c|c|c|}
\hline
Variation of input &
$f_D$-bound uncertainty&
$f_{D_s}$-bound uncertainty\\
\hline
$m_c$&$\pm\, 2.0 \%$ &$\pm 2.0 \%$ \\
\hline
$m_s$ &
-& $\pm\, 1.4 \%$ \\
\hline
$\alpha_s$& 
$\pm\, 0.7 \%$ &
$\pm\,  0.7 \%$ \\
\hline

$\langle 0 |\bar{q}q| 0\rangle$ & 
${}^{+3.5}_{-2.5}\%$ & ${}^{+2.7}_{-1.9}\%$\\
\hline
$\langle 0|\bar{s}s |0\rangle/
\langle 0| \bar{q}q |0\rangle$&
-&
$\pm\, 5 \%$\\
\hline    
d=4,5,6 condensates &
$\pm\,1.0\%$&
$\pm\,1.0\%$\\  
\hline
scale $\mu$&
${}^{+3.4}_{-2.1}\%$&
${}^{+3.6}_{-3.0}\%$\\ 
\hline
total in quadr. & 
$\pm 4.8 \%$ & $ \pm 6.9 \% $\\
\hline 
\end{tabular}
\end{center}

Summing up separate 
uncertainties in quadratures,
we obtain $\simeq \pm 10$ MeV ($\pm 20$ MeV) for
the $f_D$ ($f_{D_s}$)  bound in  (\ref{eq:1stbound}). 
The uncertainty of the $f_{D_s}$-bound is 
naturally larger,
due to the variation of $m_s$ and the spread in the 
ratio of strange and nonstrange condensates. 
Conservatively, we shift  the bounds
(\ref{eq:1stbound}) up 
by their respective total uncertainties, 
yielding our final estimate:
\be
f_{D}< 230 ~\MeV,~~ f_{D_s}< 270 ~\MeV \,.
\label{eq:bound}
\ee


\section{Discussion} 

The lattice results \cite{Follana,Aubin,
Blossier,vonHippel,Ali Khan,Heitger} for $f_{D}$ and $f_{D_s}$ 
are consistent with the upper bounds (\ref{eq:bound}). 
The decay constant of $D$-meson (\ref{eq:expD}) 
extracted from experiment also obeys 
its bound.
Formally, the experimental value (\ref{eq:expDs}) of
$f_{D_s}$ does not contradict (\ref{eq:bound}) as well, 
especially if one takes into account the 
experimental errors. 
However, it seems quite unnatural for
this bound to be almost completely 
saturated by the $D_s$ contribution. In the correlation function 
$\Pi_s(q^2)$ at timelike $q^2$  
there are intermediate hadronic states 
located above $D_s$, starting from $D^*K$
($D_s^*\pi$ is forbidden  by isospin symmetry), 
and including the radial excitations of $D_s$
and continuum hadronic states with 
$D_s$ quantum numbers.    
It is difficult to evaluate in a model-independent 
way the size of individual contributions of these states 
to the hadronic spectral density. 
Their Borel-transformed sum, a positive quantity, can be estimated,
subtracting the $D_s$-contribution from the correlation function 
$\Pi_s(M^2)$ calculated from OPE:
\be
\sum\limits_{h=D^*K,...}
\langle 0\mid j_5 |h\rangle\langle h | j_5^\dagger\!\mid 0\rangle 
e^{-m_{h}^2/M^2}=\Pi_s(M^2)-f_{D_s}^2m_{D_s}^4 e^{-m_{D_s}^2/M^2}\,,
\label{eq:sum}
\ee
and substituting the experimentally determined 
$f_{D_s}$  from (\ref{eq:expDs}). For normalization 
we divide both parts of this equation by the $D_s$ contribution
and calculate the ratio: 
\be
R_s(M^2)=\frac{\sum\limits_{h=D^*K,...}
\langle 0\mid j_5 |h\rangle\langle h | j_5^\dagger\!\mid 0\rangle 
e^{-m_{h}^2/M^2}}{f_{D_s}^2m_{D_s}^4 e^{-m_{D_s}^2/M^2}} 
\label{eq:R}
\ee
from (\ref{eq:sum}), as well as the analogous ratio 
$R(M^2)$ for the $D$-meson 
channel where, correspondingly, the experimental 
result for $f_D$ is used. 
\begin{figure}
\begin{center}
\includegraphics[width=8cm]{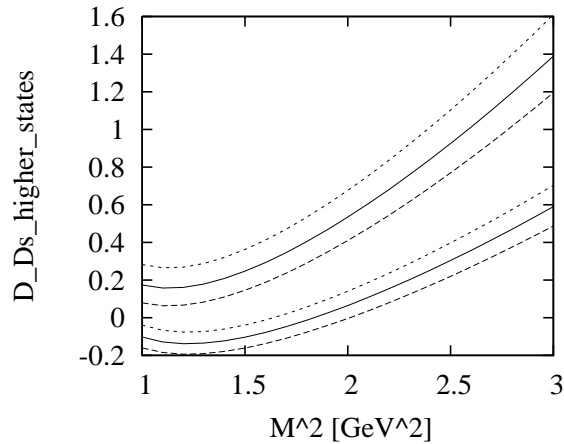}
\end{center}
\caption{\it  The lower (upper) solid curve is the sum 
of higher-state hadronic contributions
to the Borel-transformed correlation function 
in $D_s$ ($D$) channel, normalized
by the $D_s$ ($D$) contribution, as a function of the Borel
parameter squared. The 
OPE for the correlation function $\Pi_s(M^2)$ (\,$\Pi(M^2)$) 
is calculated at central values of input parameters and 
the experimental  value for $f_{D_s}$  ($f_D$) is used. Dashed
lines indicate the experimental errors in the decay constant
determination.}
\end{figure}
Note that 
in the $SU(3)_{fl}$ symmetry limit 
the correlation functions for $D_s$ and $D$ channels 
and their hadronic components are equal: 
$\Pi_s(M^2)=\Pi(M^2)$, $f_{D_s}=f_{D}$, $m_D=m_{D_s}$, $R_s(M^2)=R(M^2)$, 
etc.  Our calculation of $\Pi_s(M^2)$ and $\Pi(M^2)$ 
explicitly takes the $SU(3)_{fl}$-violation
effects into account, via 
differences of $s$- and $u,d$- quark  masses and  
$\bar{q}q$ $(q=u,d)$ and  $\bar{s}s$ condensates.
Since here we also use quite different values
of $f_{D_s}$ and $f_D$,  
the results for  the hadronic quantities $R_s(M^2)$ and $R(M^2) $, 
are  drastically different, as seen from Fig. 2.
At $1<M^2 <2$ GeV$^2$ the hadronic sum for $D_s$ is 
even negative, violating unitarity. At larger $M^2$, 
the share of higher states in the $D_s$-channel correlation
function is strongly suppressed with respect to the 
same characteristics for $D$-channel. Maximizing 
the uncertainties in OPE (as we did it in obtaining 
the conservative bounds (\ref{eq:bound})), one
can shift both curves upwards, making $R_s$ marginally 
consistent with positivity, however the large 
$SU(3)_{fl}$-difference remains.   
 A natural question is then: why is $SU(3)_{fl}$-symmetry  
so strongly violated in this correlation function?

In general, the reliability of the bounds (\ref{eq:bounds}) 
depends on the convergence of OPE. Including
the $O(\alpha_s^2)$ perturbative loops 
and condensate contributions  
up to $d=6$  we obtain a reasonable numerical convergence.
One, however, cannot completely exclude, that some 
nonperturbative, e.g., instanton-like effects contribute 
to $\Pi(M^2)$ beyond OPE, and modify the bounds. 
However, even if such effects influence 
the convergence of OPE, it is hardly possible 
that they simultaneously produce a drastic $SU(3)_{fl}$-violation.   
Moreover, a similar OPE in terms of perturbative loops and power-suppressed
local condensates for the light-quark (pseudo)scalar currents
was successfully used to obtain the lower bounds
for the light-quark masses \cite{Lellouch}, up to $O(\alpha_s^4)$ 
accuracy \cite{BCK}. 
If there were noticeable effects beyond OPE in the correlation
function of charmed pseudoscalar currents, their role 
would have become more pronounced for the light-quark mass bounds,
obtained with a light quark replacing the virtual $c$ quark.


\section{Conclusion} 

Concluding, we find that the current 
experimental result for $f_{D_s}$ only marginally 
obeys the  upper bound for this decay constant 
obtained from the OPE for the two-point correlation
function. This is in contrast to the $D$-meson case 
where the analogous bound is consistent with $f_D$ 
inferred from experiment. 
A more precise knowledge of $c$- and $s$-quark masses 
and of the ratio of strange and nonstrange quark condensates
will make the bounds more accurate.
Furthermore, the
contributions of higher states  
to the correlation function in $D_s$ channel 
estimated  using the experimental
value of $f_{D_s}$, are strongly suppressed 
as compared with the corresponding contributions for the $D$ meson
channel.  

If the  noticed difference between strange and nonstrange 
charmed meson channels remains in future, from QCD point of view, 
that could indicate a presence of unaccounted nonperturbative effects
in the pseudoscalar correlation function of charmed quarks 
which violate not only OPE, but also $SU(3)_{fl}$ symmetry. 
However such effects
seem not to manifest themselves in the lattice calculations.
   
\bigskip
{\bf Acknowledgements} 

I thank V.~Braun, K.~Chetyrkin  and M.~Jamin for useful 
discussions. 
This work is supported by the Deutsche Forschungsgemeinschaft 
under the  contract No. KH205/1-2. I also acknowledge 
the hospitality of the  Kavli Institute for Theoretical Physics China
(Beijing)  where this work was initiated.

\end{document}